\documentclass[aps,twocolumn,showpacs,preprintnumbers,amsmath,amssymb,floatfix]{revtex4}

\usepackage{graphicx}
\usepackage{dcolumn}
\usepackage{bm}
\usepackage[german,american]{babel}

\newcommand{\figref}[1]{Fig.\ \ref{#1}}

\begin{document}


\title{Backward correlations and dynamic heterogeneities: a computer study of ion dynamics}

\author{A.\ Heuer}
\email{andheuer@uni-muenster.de}

\author{M. Kunow}%
\email{kunow@uni-muenster.de}

\author{M.\ Vogel}
\email{mivogel@uni-muenster.de}

\author{R.D.\ Banhatti}
\email{banhatt@uni-muenster.de}

\affiliation{
Westf\"{a}lische Wilhelms-Universit\"{a}t M\"{u}nster\\
Institut f\"{u}r Physikalische Chemie and Sonderforschungsbereich 458\\
Schlossplatz 4/7, D-48149 M\"{u}nster, Germany
}

\date{\today}

\begin{abstract}
We analyse the correlated back and forth dynamics and dynamic
heterogeneities, i.e. the presence of fast and slow ions, for a
lithium metasilicate system via computer simulations. For this
purpose we define, in analogy to previous work in the field of
glass transition, appropriate three-time correlation functions.
They contain information about the dynamics during two successive
time intervals.  First we apply them to simple model systems in
order to clarify their information content. Afterwards we use this
formalism to analyse the lithium trajectories. A strong
back-dragging effect is observed, which also fulfills the
time-temperature superposition principle. Furthermore, it turns
out that the back-dragging effect is long-ranged and exceeds the
nearest neighbor position. In contrast, the strength of the
dynamic heterogeneities does not fulfill the time-temperature
superposition principle. The lower the temperature, the stronger
the mobility difference between fast and slow ions. The results
are then compared with the simple model systems considered here as
well as with some lattice models of ion dynamics.
\end{abstract}

\pacs{66.30.Dn}
\maketitle

\section{\label{intro}Introduction}

One basic characteristics of amorphous ion conductors is the
strong frequency dependence of the conductivity at sufficiently
low temperatures \cite{Funke,Dyre,Roling,Roling2}. According to
linear response theory, the dispersion in $\sigma(\nu)$ is
equivalent to a non-diffusive mean square displacement $\langle
r^2(t) \rangle$ and thus to the presence of correlated back and
forth jumps \cite{Funke}. Neglecting possible dynamic correlations
among adjacent ions the relation between both quantities reads
\begin{equation} \sigma(\nu) = \frac{q^2 \rho}{6 k_B T}
\int_0^\infty dt \, (d/dt)w(t) \exp(-i 2\pi \nu t)
\end{equation}
where $q$ denotes the charge and $\rho$ the density of the mobile
ions. The function $w(t)$ is defined as
\begin{equation}
w(t)= (d/dt) \langle r^2(t) \rangle.
\end{equation}
For very high frequencies one observes local dynamics which is
strongly system-dependent.  For example, in a sodium silicate
system \cite{Wong} one observes for  $\nu < 10^{11}$ Hz a
continuous decrease of $\sigma(\nu)$ with decreasing $\nu$ which
can typically be attributed to non-local dynamics \cite{Conny} and
becomes stronger for decreasing temperature. Actually, this
conclusion has been explicitly verified in recent simulations for
lithium metasilicate \cite{Heuer02}. In a double-logarithmic
representation at low temperatures the apparent exponent decreases
from one to zero until the d.c. plateau is reached, i.e.
$\sigma(\nu) = \sigma_{d.c.}$.

In general, the complexity of ion dynamics in disordered systems
is, on the one hand, related to the static disorder of the
material and, on the other hand, to the Coulomb interaction among
the ions giving rise to dynamic disorder. Therefore, one might
expect that both the static as well as the dynamic disorder
enhances the number of correlated back and forth jumps (see,
however, Ref.\cite{Reinisch}) although in recent simulations of
sodium silicate at $T = 2000$ K no such correlations have been
observed \cite{Jund02}. Furthermore, the presence of different
environments of the network might imply that at a given time some
ions are more mobile than other ions, i.e. there exist dynamic
heterogeneities. Unfortunately, no direct information about
dynamic heterogeneities can be gained from conductivity
experiments or, equivalently, from the time dependence of the mean
square displacement. Some progress has been achieved on the basis
of simulations. Comparing the mean square displacement of
different ions during a given time a broad distribution has been
observed \cite{Habasaki02}. This basically corresponds to the
presence of a large non-gaussian parameter which is known to
represent dynamic heterogeneities quite well \cite{Doliwa99}.
Other groups found an interesting spatial structure of the mobile
regions \cite{Kob1,Horbach3}.

Another system with intrinsic complex dynamics is a glass-forming
liquid. A tagged particle in a glass-former experiences a dramatic
slowing down with decreasing temperature \cite{Angell}. To a large
extent this is related to the cage effect since at low
temperatures the neighbour particles have a strong confining
effect on the central particle \cite{Goetze}. This gives rise to
strongly subdiffusive dynamics, i.e. to correlated back and forth
dynamics. Furthermore glass-forming systems display dynamic
heterogeneities \cite{Spiess,Heuer95,Sillescu96,Ediger}.  This
property has been quantified by invoking appropriate three-time
correlation functions, containing information about the dynamics
during two subsequent time intervals \cite{Okun,Doliwa98,Qian}.

Using these three-time correlation functions one can hope to
answer several basic questions about the complexity of ion
dynamics. How relevant are backward correlations? Are back and
forth correlations restricted to nearest-neighbor ionic positions?
How does the tendency of these back and forth correlations depend
on the time scale of investigation? Do dynamic heterogeneities
depend on temperature and thus invalidate the time-temperature
superposition principle, observed for many other quantities like
the conductivity? This and other aspects of ionic dynamics will be
analysed in this paper by invoking appropriate three-time
correlation functions. They will be applied to computer generated
trajectories of lithium metasilicate (Li$_2$O)(SiO$_2$). Very
recently, first measurements of three-time correlations have been
conducted via multidimensional NMR. These results clearly showed
that dynamic heterogeneities are omnipresent in disordered ionic
conductors \cite{NMR}

The organization of this paper is as follows. In Section 2 we
describe the technical aspects of the simulation and discuss the
numerical tools. Section 3 introduces the concept of three-time
correlations. Section 4 contains the analysis of simple model
systems for which we clarify the information content of the
three-time correlation functions. The results of our simulations as
well as their interpretation are presented in Section 5. We close
with a discussion and a summary in Section 6.

\section{Simulation}

The potential energy for the lithium silicate system is chosen to
be the sum of a Buckingham and a Coulomb pair potential ($i,j$
denote the species lithium, oxygen, or silicon, respectively)
\begin{equation}
U_{ij}(r) = \frac{q_i q_j e^2}{r} - \frac{C_{ij}}{r^6} + A_{ij}
\exp(-B_{ij} r ).
\end{equation}
The parameters have been determined by Habasaki et al.
\cite{Habasaki2,Habasaki1}. Details of our simulation can be found
in \cite{Banhatti,Heuer02}. Summarizing, we performed molecular
dynamics simulations with a time step of 2 fs and a density of
$\rho = 2.34$ g cm$^{-3}$.  Periodic boundary conditions were
used. The system size is 1152 particles, thus containing 384
lithium ions. The trajectories were generated by an appropriately
modified version of the  MOLDY software package, supplied by K.
Refson \cite{Refson}. The length of the production runs was 16 ns
 after an equilibration time of ca.\ 10 ns at the two lowest temperatures.
Even at $T$ = 640 K the mean square displacement during the production
time was larger than 60 \AA$^2$ (nearest-neighbour Li-Li distance 2.6 \AA);
see Ref.\ \cite{Heuer02}. All configurations were equilibrated at
T = 1500 K. The computer glass transition is approximately 1100 K
\cite{Heuer02,Banhatti}. In this work we present simulations for
five temperatures (T = 1240 K, T = 980 K, 750 K, 700 K, and 640
K). For all temperatures, the lithium subsystem has been
equilibrated before starting the production run. In particular at
the lower temperatures the network fluctuations are very small
(mean square displacement of 0.3 \AA$^2$).

\section{Three-time correlations}

\subsection{Definition of three-time correlations}

\begin{figure}[t]
  \includegraphics[width=8.6cm]{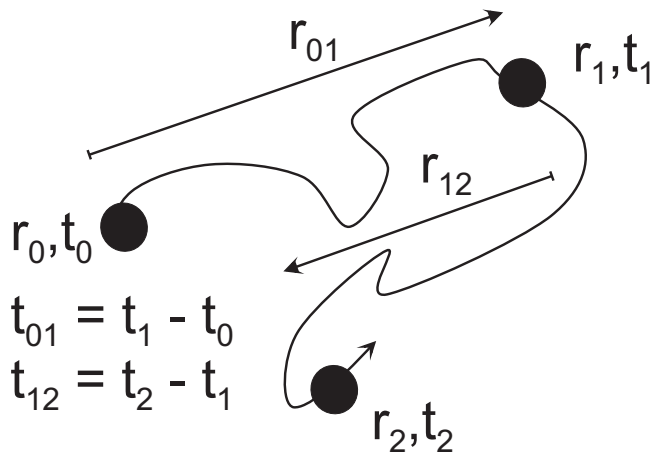}
  \caption{\label{fig_1}Sketch of a single-particle dynamics in order to
clarify the definition of $r_{01}$ and $r_{12}$.}
\end{figure}

As mentioned above, the dispersive behavior of the conductivity
or, equivalently, the non-diffusive behavior of the mean square
displacement, has been related to correlated back and forth
dynamics. Here we want to introduce a formalism which allows one
to elucidate the properties of back and forth dynamics in detail.
We remind the reader that standard observables for the
characterization of dynamical properties like the mean square
displacement correlate the position of individual ions at two
successive times $t_0$ and $t_1$. For a calculation of $\langle
r^2(t) \rangle $ one averages over the configurations at all times
$t_0$ and $t_1$ such that $t_1 - t_0 = t$. Back and forth
dynamics, however, is related to the properties of particles
during two subsequent time intervals. Formally, this can be
described as a three-time correlation, including a third time $t_2
> t_1$. The idea is sketched in \figref{fig_1}. $r_{01}$ denotes the
distance a particle moves during the first time interval of length
$t_{01} = t_1 - t_0$. The value of $r_{12}$ denotes the motion
during the second time interval of length $t_{12} = t_2 - t_1$ as
projected on the direction of the motion during the first time
interval. In case of a backjump, as shown for the example in
\figref{fig_1}, the value of $r_{12}$ is counted negative. The additional
information about the dynamics during two successive time
intervals as compared to the dynamics during a single time
interval is contained in the conditional probability function
$p(r_{12}|r_{01})$ which denotes the probability for a specific
value of $r_{12}$ under the condition that the particle has moved
the distance $r_{01}$ in the first time interval.

\subsection{Moments}

Rather than analysing the full probability function, we
concentrate on the first moment $\bar{r}(r_{01})$ and the second
moment $v(r_{01})\equiv \langle (r_{12} - \bar{r}(r_{01}))^2
(r_{01})\rangle$. The interpretation of both functions is
straightforward. $\bar{r}(r_{01})$ contains information about the
relevance of back and forth dynamics. In case that the direction
of the dynamics during two successive time intervals is
uncorrelated one expects $ \bar{r}(r_{01}) = 0$. In contrast, a
negative value of $\bar{r}$ is direct evidence of the presence of
back and forth dynamics.

The second moment $ v (r_{01})$ yields information about the
presence of dynamic heterogeneities. In case that all particles
have the same mobility, the distance moved in the second time
interval is independent of the distance moved in the first time
interval. Thus $v(r_{01})$ would {\it not} depend on $r_{01}$. A
dependence on $r_{01}$ will be observed if there exist fast and
slow ions. The subensemble of ions with small $r_{01}$ will
preferably contain slow ions whereas for ions with large $r_{01}$
it is vice versa. Therefore, ions with small $r_{01}$ will on
average move less in the second time interval $t_{12}$ than ions
with large $r_{01}$, resulting in a monotonous increase of
$v(r_{01})$ with $r_{01}$.

\subsection{First moment in the limit $t_{01} \rightarrow 0$}

It is possible to establish a direct relation between the mean
square displacement and the first moment $\bar{r}(r_{01})$ for the
case of a discrete hopping model in the limit $t_{01} \rightarrow
0$. For reasons of simplicity we take a 1D model with distances
$d_0$ between the individual sites. One may start with the simple
relation
\begin{equation}
\label{r2}
\langle r^2(t_{01} + t_{12}) \rangle = \langle
r^2(t_{01}) \rangle + 2 \langle r(t_{01}) r(t_{12}) \rangle +
\langle r^2(t_{12}) \rangle
\end{equation}
which is valid for stationary processes. On the left side one can
perform a linear expansion around $t_{01}=0$. On the right  side
one may use the fact that for very small $t_{01}$ the term $\langle
r^2(t_{01}) \rangle $ can be written as $\Gamma_{eff}d_0^2 t_{01}$
(corresponding to the short-time diffusion in pure hopping models
with an effective escape rate $\Gamma_{eff}$). Since for very short
times the system can only jump to the nearest neighbor site the
term $\langle r(t_{01}) r(t_{12}) \rangle$ can be expressed as
$\Gamma_{eff} d_0 t_{01}
\bar{r}(d_0)$. Inserting these relations into Eq.\ref{r2} one
finally ends up with
\begin{equation}
\label{firstmoment}
 \frac{\bar{r}(d_0)}{d_0} = \frac{1}{2} \left [
\frac{w(t_{12})}{w(0)} - 1 \right ].
\end{equation}

This relation directly shows that exactly in case of diffusive
dynamics, i.e. $w(t) = const$,  one has $\bar{r}(d_0)= 0$, i.e. no
correlated back and forth dynamics. For subdiffusive behavior one
obtains (beyond a possible oscillatory regime of the mean square
displacement)  $ 0 < w(t) < w(0)$ and thus $-1/2 <
\bar{r}(d_0)/d_0 < 0$. Thus subdiffusive behavior is
equivalent to the presence of correlated forward and backward
jumps. Furthermore validity of Eq.\ref{firstmoment} implies that
the first moment $ \bar{r}(d_0)$ has a lower limit $-(1/2)d_0$
which is reached if $w(t) = 0$, i.e. for the long-time limit of
localized dynamics where any forth jump is followed by a back jump
. Finally one can see that the range of dispersion, i.e.
$w(t\rightarrow \infty) / w(0)$, can be related to the first moment
$ \bar{r}(d_0)$ in the limit $t_{01} \rightarrow 0$ and $t_{12}
\rightarrow \infty$.

This formal treatment has been performed for a hopping model with
discrete sites. This may be considered as an appropriate model
also for more realistic systems for which the particles will
fluctuate around the individual sites. In particular this is the
case for the ion conductor, studied in this work (see below for
more details). In contrast, for glass-forming systems hopping
dynamics is not so relevant. Therefore it is not possible to
formulate such a simple relation between the derivative $w(t)$ of
the mean square displacement and the first moment $
\bar{r}(d_0)$.

Eq.\ref{firstmoment} also implies that the first moment
$\bar{r}(d_0)$ in the limit $t_{01} \rightarrow 0$ does {\it not}
contain new information as compared to the mean square
displacement. For finite $t_{01}$, however, the first moment can
no longer be predicted from $w(t)$. Thus new information as
compared to the mean square displacement about the nature of
correlated back and forth dynamics becomes available. Of
particular interest is the dependence of the first moment on
$r_{01}$. For example one can learn whether there exists a
long-range back-dragging effect, as implied e.g. in the
percolation approach of ion dynamics, or whether after jumping to
the next nearest neighbor site the memory about the initial site
has been basically wiped out.

\section{Model calculations}

In order to clarify the information content of the first moment
$\bar{r}(r_{01})$ and the second moment  $v(r_{01})$ we first
calculate them for some simple one-dimensional models. We always
consider the case of stochastic dynamics since we are interested
in time scales which are beyond the ballistic regime.

\subsection{Harmonic oscillator}

This model is relevant to describe the backward and forward
dynamics in the individual potential wells. We consider a harmonic
oscillator with minimum at $r=0$. Both times $t_{01}, t_{12}$ are
longer than the equilibration time of the oscillator, i.e. the mean
square displacement is already constant. We first calculate the
probability $p_1(r_1)$ that after the first time interval the
system is at $r_1$  after a motion of $r_{01}$ to the right. This
can be formally calculated as
\begin{equation}
\label{eqp} p_1(r_1) = \int_0^\infty dr_0 \,  p_0(r_0) p(r_1|r_0)
\delta (r_{01} - (r_1 - r_0))
\end{equation}
where $p_0(r_0)$ denotes the equilibrium distribution and
$p(r_1|r_0)$ is the  probability to move from $r_0$ to $r_1$
during time $t_{01}$. For large $t_{01}$ the latter term is
identical to the equilibrium distribution, i.e. $p_0(r_1)$. The
resulting gaussian integral can be easily solved. Here we are
particularly interested in the first moment $\langle r_1 \rangle$
of $p_1(r_1)$. One obtains
\begin{equation}
 \langle{r}_1 \rangle = (1/2) r_{01}.
\end{equation}
For long $t_{12}$ the system acquires the average $\langle r_2
\rangle = 0$, yielding
\begin{equation}
\bar{r}(r_{01}) = \langle r_2 \rangle  -  \langle r_1 \rangle
= -(1/2) r_{01}.
\end{equation}
Thus the back-dragging effect in the second time interval is
proportional to the distance moved in the first time interval.

\subsection{Periodic potential}

\begin{figure}[t]
  \includegraphics[width=8.6cm]{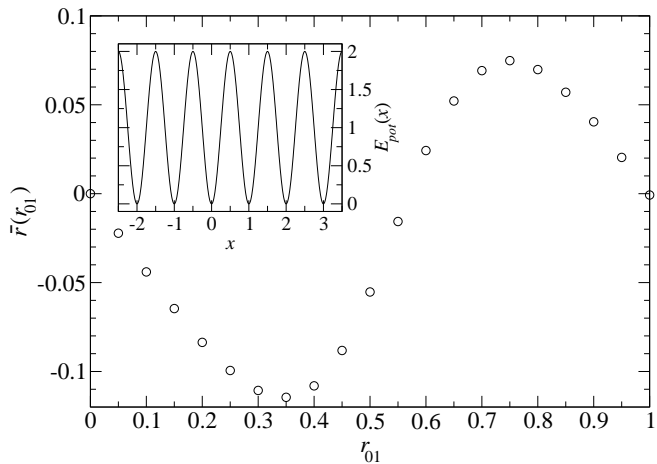}
  \caption{\label{fig_2}$\bar{r}(r_{01})$ as obtained from Monte
Carlo simulations of the simple periodic cos-potential. The
potential is shown in the inset.}
\end{figure}

In the next step, we analyse a potential which involves local
vibrations as well as hopping dynamics. Here we consider a
periodic potential $E_{pot}(r)$ with minima at integer values of
$r$, defined as
\begin{equation}
V(r) = V_0 [1-\cos(2\pi r)].
\end{equation}
We are interested in the stochastic dynamics of a particle. In
general, one has to resort to numerical simulations to calculate
$\bar{r}(r_{01})$. Here we have modelled the dynamics via standard
kinetic Monte Carlo simulations at the temperature $T = 0.4 V_0 $
with step sizes much smaller than the distance of two adjacent
minima. The qualitative features of the result do not depend on
the exact value of this temperature. We have chosen $t_{01}$ such
that on this time scale a particle leaves the initial well with a
probability of approximately 50\%. Furthermore we have chosen
$t_{12} = 10 \, t_{01}$. The result of this simulation is shown in
\figref{fig_2}. The dependence of $\bar{r}(r_{01})$ on $r_{01}$ can be
understood from simple arguments:
 a) For $r_{01} \ll 1$, one
  basically has the behavior of a harmonic oscillator. b) For $r_{01}
= 0.5$, the particle has
 typically moved to a position close to the saddle between two
 wells.  For infinite $t_{01}$, simple symmetry
considerations show
 that $\langle r_1 \rangle $ is exactly on the saddle.  Since a
 particle on a saddle does
 not experience any effective net force to any
 side one has $\bar{r}(0.5) = 0$.  For finite $t_{01}$, a motion of
 $r_{01} = 0.5$ on average leaves the particle in the initial well \cite{expl1}.
 Thus the
function $\bar{r}(r_{01})$ has its zero for $r_{01}$ slightly
larger than 0.5. c) For $r_{01}$ approaching 1, the particle has
definitely crossed a saddle. Now the effective force points in the
same direction as the initial jump direction. This results in a
positive value of $\bar{r}(r_{01})$. d) For $r_{01} = 1$, one has
the same result as for $r_{01} = 0$, i.e. $\bar{r}(r_{01}) = 0$.

\subsection{Potential with alternating barriers}

\begin{figure}[t]
  \includegraphics[width=8.6cm]{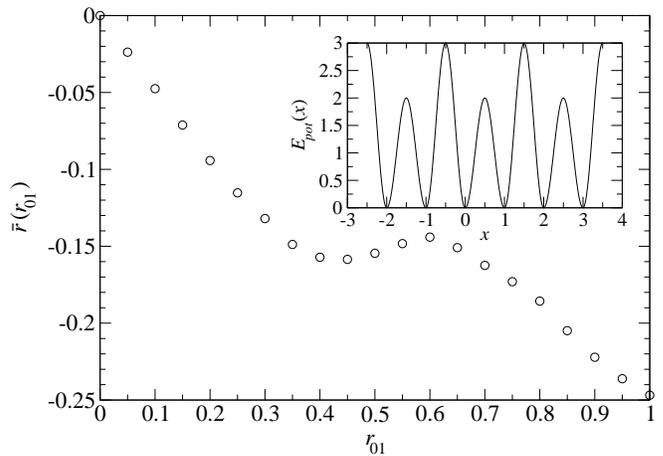}
  \caption{\label{fig_3}$\bar{r}(r_{01})$ as obtained from Monte
Carlo simulations of a periodic potential with alternating
barriers. The potential is shown in the inset.}
\end{figure}

So far we have only discussed backwards dynamics due to intrawell
dynamics. For an ion conductor one expects that dynamic forward
backward correlations either result from static or from dynamic
disorder. Here we briefly discuss a very simple model which
contains non-trivial forward backward dynamics. It is shown in the
inset of \figref{fig_3}. Due to the alternating barrier heights a particle
performs several forward and backward jumps until it can escape
from the local cage. We denote the two different transition rates
by $\Gamma_1$ and $\Gamma_2$ ($\Gamma_1 \gg \Gamma_2$). Our goal
is to calculate $\bar{r}(r_{01} = 1)$. For very long $t_{01}$,
i.e. $t_{01} \Gamma_2 \gg 1$ the dynamics resembles that of a
random walk so that forward backward correlations should not be
relevant, i.e. $\bar{r}(1)\approx 0$. Therefore we restrict
ourselves to the case $t_{01} \Gamma_2 \ll 1$.  The analysis is
presented in Appendix \ref{app1}.

We obtain
\begin{eqnarray}
\label{eqshort}
\bar{r}(1) &=&-(1/2) (1 - 2\Gamma_2/\Gamma_1)(1 - 2\Gamma_2/\Gamma_1) \\
&& *(1 - \exp(-2\Gamma_1 t_{12})) \qquad\qquad t_{01} \Gamma_1 \ll 1 \nonumber\\
\label{eqlong}
\bar{r}(1) &=&-(1/2) (1 -2\Gamma_2/\Gamma_1)(1 - \Gamma_2 t_{01}) \\
&&*(1 - \exp(-2\Gamma_1 t_{12}))  \qquad\qquad t_{01} \Gamma_1 \gg 1. \nonumber
\end{eqnarray}
Thus there exists a limiting value for $t_{01} \rightarrow 0$
which is reached for $t_{01} \approx 1/\Gamma_1$, i.e. the time
scale of the fastest jump process. For larger values of $t_{01}$
the back-dragging effect decreases with time; see Eq.\ref{eqlong}.
$\bar{r}(1)$ approaches zero for $t_{01}$ of the order of
$1/\Gamma_2$. The physical reason is that for longer times
$t_{01}$ the particle may also cross the high barrier during the
first time interval. These events strongly reduce the total
back-dragging effect in the subsequent time interval. Thus the
$t_{01}$-dependence contains valuable information about the
time-scales involved in the dynamics and indicates at which time
scale (here: $t_{01} \approx 1/\Gamma_2$)  a simple random-walk
description becomes relevant.

For $t_{12} \rightarrow 0$ one obtains  $\bar{r}\rightarrow 0$.
This limit is trivial since there is no dynamics during the second
time interval. In contrast, for $t_{12}
\rightarrow
\infty$ the backjump effect is largest. Thus it is this limit which
is relevant to judge the maximum backjump capabilities. The rest of
the discussion for this model system deals with this case.

 To show the full dependence of
$\bar{r}(r_{01})$ on $r_{01}$ we again performed Monte Carlo
simulations for a potential with alternating barriers. This
potential was generated from the cos-potential, discussed above,
by scaling the cos-potential by a factor 1.5 in the intervals
$..., [-1,0], [1,2], [3,4], ...$. The temperature was $T =
0.5V_0$. $t_{01}$ is chosen such that the short-time  limit
$t_{01} < 1/\Gamma_1$ is fulfilled whereas $t_{12}$ corresponds to
the long-time limit $t_{12} \gg 1/\Gamma_2$. The transition rate
is proportional to the attempt frequency in the minimum, which
scales with the square root of the force constant and with the
Boltzmann factor. Thus one expects $\Gamma_2/\Gamma_1 \approx
\sqrt{1.5} \exp((2-3)/0.5) \approx 0.17$ which, according to
Eq.\ref{eqshort}, yields $\bar{r}(1) \approx 0.22$.

 The result for $\bar{r}(r_{01})$ is shown in
\figref{fig_3}. It resembles that of a periodic potential except for a
systematic downward trend. Thus we have a superposition of the
correlation effects of simple periodic potentials with wells and
barriers and of barriers with different heights.  Actually, it
turns out that our estimate of $\bar{r}(1)$ agrees reasonably well
with the numerical value of ca. 0.25.

\subsection{Random trap model}

\begin{figure}[t]
  \includegraphics[width=8.6cm]{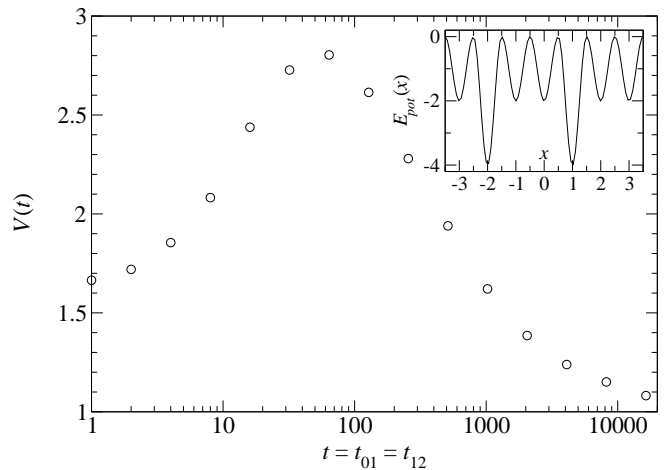}
  \caption{\label{fig_4}$V(t)$ as obtained from Monte
Carlo simulations of a potential with random traps. The potential
is shown in the inset.}
\end{figure}

Finally, we want to present a very simple model which allows one
to grasp the relevant features of the second moment $v(r_{01})$.
We start with an (possibly high-dimensional) array of traps with
different depths. Such models have been extensively studied in the
context of supercooled liquids \cite{Bouchaud,Odagaki,Doliwa02}.
Different rates $\Gamma_i$ are randomly attributed to the
different traps. For such a simple model the dynamics is purely
diffusive. Here we are specifically interested in the ratio $V(t)
\equiv  v(r_{01} = 1,t)/v(r_{01} = 0,t)$ for $t_{01} = t_{12}=t$.
According to our discussion in Section III  $V(t)$ is a measure
for the relevance of dynamic heterogeneities.

As shown in Appendix \ref{app2} one can derive the relation
\begin{equation}
V(t \rightarrow 0) =
 \langle \Gamma \rangle\left \langle \frac{1}{\Gamma} \right \rangle.
\end{equation}
 Without dynamic heterogeneities it does
not matter whether or not a particle moves in the first time
interval such that $v(r_{01} = 1,t) = v(r_{01} = 0,t)$. In case of
no dynamic heterogeneities, i.e. a single value of $\Gamma$, one
trivially has $\langle \Gamma \rangle = 1/\langle 1/\Gamma
\rangle$ and thus $V(t) = 1$. For a distribution of jump rates the
product $\langle \Gamma \rangle \langle 1/\Gamma \rangle $ is
larger than one. This can be easily rationalized for a bimodal rate
distribution with rates $\Gamma_1$, $\Gamma_2$ and weights $a_1$,
$a_2 $, respectively. A straightforward calculation yields $V(t) =
1 + (a_1 a_2)(\Gamma_1 - \Gamma_2)^2/(\Gamma_1 \Gamma_2)$ which for
a bimodal distribution is strictly larger than one. For stronger
dynamic heterogeneities, i.e. a broader distribution of rates
$\Gamma$, $V(t)$ also increases.  Thus $V(t \rightarrow 0)$ is a
direct measure for dynamic heterogeneities.

In the opposite limit $t \rightarrow \infty$, a particle with
$r_{01} = 0$ has by chance returned to the original position. This
implies that the  condition $r_{01} = 0$ no longer implies any
dynamical selection of slow particles. Thus one expects $V(t
\rightarrow \infty) = 1$. Whereas the detailed time-dependence of
$V(t)$ depends on more details of the model like the number of
neighbor traps, the limiting values are generally valid.

In order to visualize the full time-dependence, and to check our
analytic expression we have performed Monte-Carlo simulations for
a one-dimensional random trap model. We have chosen two escape
rates, characterized by $a_1 = 0.035, a_2 = 0.965, \Gamma_1 =
0.005, \Gamma_2 = 0.1$. For this specific choice of parameters one
has $\langle \Gamma \rangle \langle 1/\Gamma \rangle = 1.61$. The
time dependence of $V(t)$ is shown in \figref{fig_4} as obtained via Monte
Carlo simulations. For this simple model the algorithm can be
implemented in a straightforward way. One observes that the
theoretical short-time and long-time limits are confirmed by the
numerical data. Interestingly, $V(t)$ displays a maximum.
Qualitatively, this means that at the time scale of the maximum
the effects of dynamic heterogeneities are most pronounced. A more
detailed discussion of the time-dependence and of the maximum is
beyond the scope of the present paper.

\section{Results}

\subsection{Previous results}

\begin{figure}[t]
  \includegraphics[width=8.6cm]{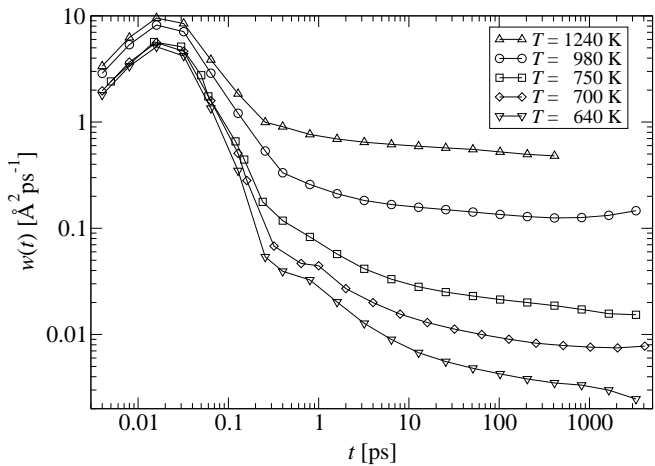}
  \caption{\label{fig_5}The derivative $w(t)$ of the mean square displacement shown for
different temperatures. Note that for $t > 1 ps$ the function
$w(t)$ is mainly governed by long-range dynamics (see text).}
\end{figure}

\begin{figure}[t]
  \includegraphics[width=8.6cm]{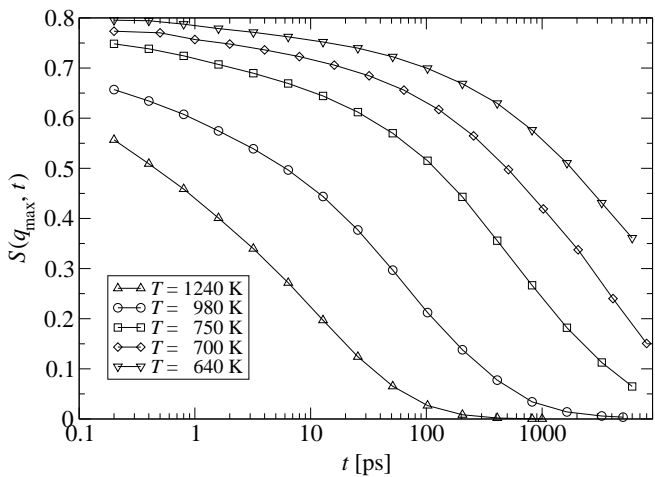}
  \caption{\label{fig_6}The incoherent scattering function $S(q_{max},t)$ at different
temperatures. The solid lines correspond to KWW-fits with $\beta=0.45$.}
\end{figure}

It has been observed that lithium trajectories can be described as
a series of local vibrations and jumps between adjacent ionic
sites; see e.g. Refs.\cite{Kieffer1,Smith} for similar features in
previous simulations on alkali silicates. In our recent work we
have shown that the van Hove self correlation function displays a
strong peak for $d_0 \approx 2.6$ \AA. This peak is separated by a
minimum at 1.5 \AA \, from the peak at the origin. The
interpretation is straightforward. The potential energy landscape,
as supplied by the network, provides lithium sites with an average
distance of 2.6 \AA \, which are separated by a saddle.

Furthermore, it turned out that the mean square displacement
curves at different temperatures show time-temperature
superposition. In \figref{fig_5} we show their derivative, i.e. $w(t)$, for
the five temperatures analysed in our prior work \cite{Heuer02}.
As shown in \cite{Heuer02} the function $w(t)$ for $t
> 1 ps$ is due to processes which involve long-range dynamical
processes ($|\vec{r}(t) - \vec{r}(0)| \ge$ 1.5 \AA). Whether or
not these processes can always be interpreted as jumps is
currently under investigation. For $t < 1 ps$ the function $w(t)$
is dominated by localized processes of the lithium ions.  Since
the presence of back and forth dynamics is equivalent to a
decrease of $w(t)$ with time, one directly sees that at least in
the case of the two lowest temperatures $T = 700$K and $T = 640$K
long-range back and forth correlations are indeed important. In
order to characterize the typical time scales of the long-range
dynamical processes we show in \figref{fig_6} the incoherent scattering
function $S(q_{max},t)$ where $q_{max} = 2\pi/d_0$. It is a
measure of the probability that an ion is still (or again) at the
initial site after time t. All data can be consistently fitted
with a KWW-function $f(t) \propto \exp(-(t/\tau)^\beta)$ with
$\beta = 0.45$.

\subsection{Back and forth dynamics}

\begin{figure}[t]
  \includegraphics[width=8.6cm]{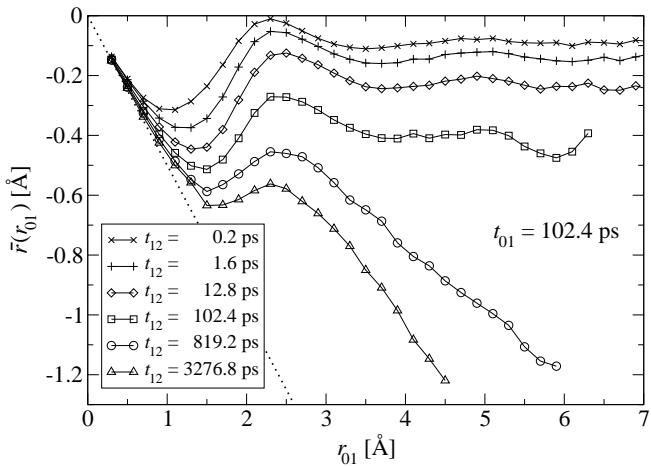}
  \caption{\label{fig_7}The first moment $\bar{r}(r_{01})$ for $t_{01} = 102.4 ps$ and for
different $t_{12}$ at $T= 750$ K. The broken line corresponds to
$\bar{r}(r_{01}) = -(1/2) _{01}$.}
\end{figure}

\begin{figure}[t]
  \includegraphics[width=8.6cm]{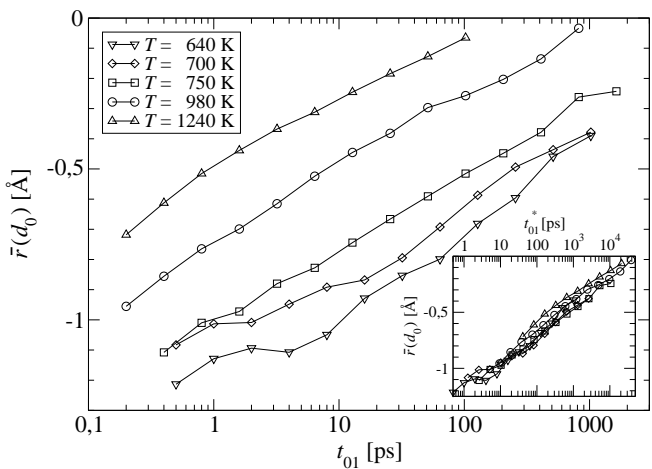}
  \caption{\label{fig_8}The first moment $\bar{r}(d_0 = 2.6 $\AA \,) in dependence of
$t_{01}$. The value of $t_{12}$ has been scaled for every
temperature on the basis of the diffusion constant [$t_{12}$(T =
640 K) = 3276.8 ps]. In the inset also the values of $t_{01}$ are
scaled by the diffusion constant such that $t_{01} (T = 640K)
\equiv t^*_{01} (T = 640K)$.}
\end{figure}

A quantitative analysis of the back and forth dynamics is possible
on the basis of the first moment $\bar{r}(r_{01})$.
 In \figref{fig_7}, we have plotted this function for a fixed value of $t_{01}=102.4$ ps
 and different $t_{12}$ at $T = 750$ K. For large $r_{01}$ the statistics
 of these curves becomes quite poor since only a few ions
 participate. This problem becomes worse for short $t_{01}$ and/or
 long $t_{12}$. In this and the following plot we restrict
 ourselves to the
 $r_{01}$-regions which possess a reasonable signal-to-noise ratio as
 estimated from the fluctuations of the curves.

 On a qualitative level the
 dependence on $r_{01}$ resembles that shown in \figref{fig_3}.
For small $r_{01}$ one recovers the harmonic behavior, as seen from
the very good agreement with the broken line $\bar{r}(r_{01}) =
-(1/2) r_{01}$. Furthermore the non-monotonic behavior directly
reflects the presence of a saddle between adjacent lithium sites.
\figref{fig_7} clearly reveals that the fraction of back and forth dynamics
increases with increasing $t_{12}$. This is expected from our
theoretical considerations; see above.

Interestingly, for the two largest values of $t_{12}$ the function
$\bar{r}(r_{01})$ decays further for $r_{01} > d_0 = 2.6$ \AA.
Thus back-dragging effects become stronger when jumping into the
second nearest neighbor shell during $t_{01}$. This observation
already goes beyond a scenario which is only based on back and
forth correlations between adjacent lithium sites.

Of particular interest is the dependence on $t_{01}$ as already
discussed for the alternate barrier model. Since we are mainly
interested in correlated back and forth dynamics between
nearest-neighbor positions we focus on the value of $\bar{r}(d_0)$
which characterizes the subsequent dynamics of a particle which has
jumped to the nearest neighbor distance in the first time interval.
We choose a large but constant value for $t_{12}$ (4.1 ns ps for T
= 640 K) and vary $t_{01}$. Comparison of different temperatures is
achieved by choosing the respective value of $t_{12}$ approximately
proportional to the inverse diffusion constant. This results in
$t_{12} = 26 ps (T = 1240 K), t_{12} = 102 ps (T = 980 K), t_{12} =
0.8 ns (T = 750 K), t_{12} = 2.0 ns (T = 700 K), t_{12} = 4.1 ns (T
= 640 K)$. The data are shown in \figref{fig_8}. Obviously, for longer times
$t_{01}$ the back-dragging effect becomes much smaller. This agrees
with the theoretical considerations, discussed for the alternate
barrier model. The limiting value $\bar{r}(d_0) = -1.3$ \AA \, is
only approximately reached for the lowest temperature $T = 640$ K.
Interestingly, the dependence on $t_{01}$ is very gradual and
extends over several decades of time. This shows that there exists
a broad distribution of barriers experienced by the lithium ions.

For a fixed value of $t_{01}$ the relevance of correlated back and
forth dynamics increases with decreasing temperature.  From our
previous discussion the short-time limit of $t_{01} \approx 1 $ ps
is related to the dispersion according to Eq. \ref{firstmoment}.
Since only for $t_{01} > 1$ ps long-range processes become
relevant one should choose $t_{01} = 1$ ps as the short-time limit
in Eq.\ref{firstmoment}.  Checking it, e.g. for $T = 750$K, one
obtains $\bar{r}(d_0) = -1$ \AA \, which according to Eq.
\ref{firstmoment} corresponds to $w(t\rightarrow \infty)/w(1ps) =
0.23$. This agrees with the measured value of ca. 0.2 (see \figref{fig_5}).
Note that Eq.\ref{firstmoment} is based on a strict hopping
picture. Therefore one would not expect an exact agreement between
both values. In any event, since only for $T \le 750$ K
$\bar{r}(d_0)$ comes close to the limiting value of $-1.3$ \AA \,
the back-dragging effect and, equivalently, the dispersion in the
mean square displacement are relevant only in this low-temperature
range.

In the inset of \figref{fig_8} the individual curves are scaled with the
scaling factor $D(T)/D(T=640K)$, thereby introducing the scaled
time $t^*_{01}$. Within the statistical noise one can see a decent
superposition, in particular for the three lowest temperatures.
Thus one observes a time-temperature superposition also for this
rather involved quantity of correlated back and forth dynamics.

\subsection{Dynamic heterogeneities}

\begin{figure}[t]
  \includegraphics[width=8.6cm]{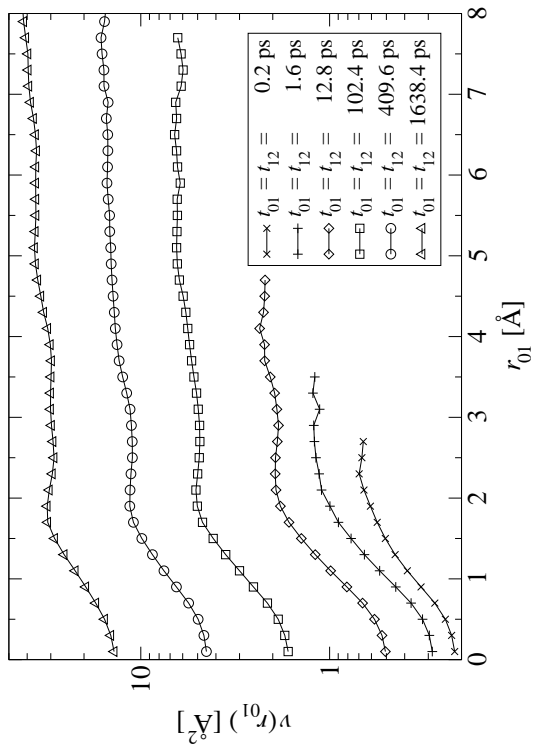}
  \caption{\label{fig_9}$v(r_{01})$  at $T = 750$K for different
choices of $t_{01}=t_{12}$.}
\end{figure}

\begin{figure}[t]
  \includegraphics[width=8.6cm]{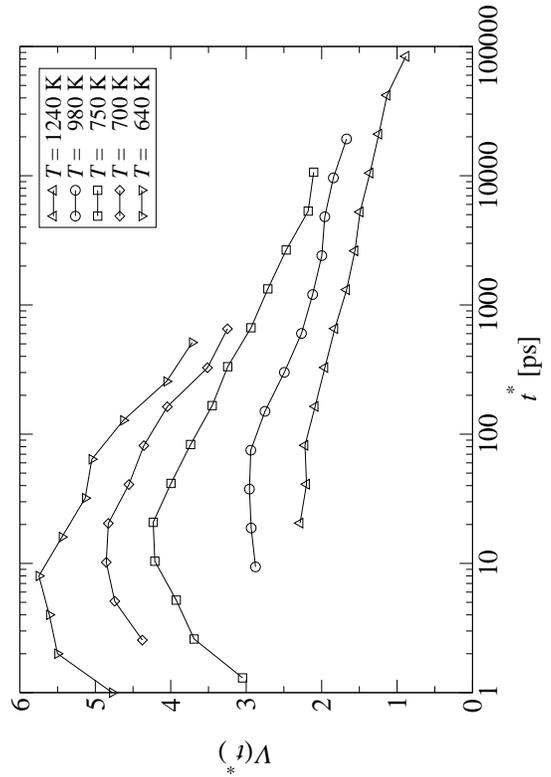}
  \caption{\label{fig_10}$V(t^*)= v(d_0)/v(0)$  at different temperatures using scaled times.}
\end{figure}

Finally we present results for the second moment $v(r_{01})$.
 In \figref{fig_9}
we have plotted $v(r_{01})$ for $T = 750$ K and different values
for  $t_{01} = t_{12}$. One can clearly distinguish two regimes
which are approximately separated by 2 \AA. For $r_{01} \ll$ 2\AA
\,,  the second moment $v(r_{01})$ is significantly smaller than
for values  $r_{01} \ge 2$ \AA. As discussed above this is a clear
signature of dynamic heterogeneities. After a jump of 2 \AA \, an
ion has  basically achieved to cross the saddle and is already
part of the adjacent ionic site. This means that a particle which
has crossed the local saddle during the first time interval is
much faster during the subsequent time interval. Starting from
$r_{01} = d_0$ the second moment $v(r_{01})$ further increases
with increasing $r_{01}$. This further increase, however, is much
weaker. This observation would be compatible with only minor
spatial correlations among ionic sites of similar mobility. In
particular, it would contradict the scenario of compact regions of
ionic sites each with a different ionic mobility. In this case, a
particle which has jumped twice would, on average, belong to a
faster region than particles which only jumped once during the
first time interval. This difference would show up in the mean
square displacement during the second time interval since
particles in the faster region would, on average, also jump
further in the second time interval. This would lead to a strong
increase of $v(r_{01})$ beyond the nearest neighbor distance
$d_0$.

In order to study the temperature dependence and the
time-dependence in greater detail we have calculated $V(t) \equiv
v(d_0)/v(0)$ in dependence of $t = t_{01} = t_{12}$ and for
different temperatures. $V(t)$ is a direct measure for the degree
of dynamic heterogeneity on the length scale of the
nearest-neighbor distance. In order to compare the different
temperatures we have again scaled all times by the ratio
$D(T)/D(T=640K)$. The results are shown in \figref{fig_10}. One can clearly
see that the degree of heterogeneity changes with temperature. The
lower the temperature the stronger the dynamic heterogeneity. Thus
on the level of dynamic heterogeneities the time-temperature
superposition principle does not hold.  For the lowest temperature
$V(t)$ is approximately 5.5 at the maximum. This number directly
implies that the mean square displacement of a particle (corrected
for possible backwards correlations) is 5.5 times larger in the
second time interval if it has performed a jump in the first time
interval as compared to particles which are still at the initial
site after $t_{01}$.

\section{Discussion}

In this paper we have shown how analysis of three-time
correlations can be used to get model-free information about the
nature of the complex ion dynamics. The characteristics of back
and forth dynamics is reflected by the first moment of the
three-time conditional probability function, the dynamic
heterogeneities by the second moment. The main results are: (i)
the long-range backward correlations beyond the nearest neighbor
position, (ii) the gradual decrease of backward correlations with
increasing $t_{01}$, (iii) the time-temperature superposition
principle for correlated back and forth dynamics, (iv) the
significant dynamic heterogeneities at low temperatures, and (v)
the lack of time-temperature superposition for the dynamic
heterogeneities. In what follows we discuss these results in more
detail.

For ion conductors at low temperatures the basically immobile
network serves as a pseudo-external field for the lithium ions.
Therefore it has been attempted to model the lithium dynamics by
simple lattice models like the random barrier or the random energy
model \cite{Dyre,Marburg}.  Also in these models the long-range
backward correlations are present. Only in simple models like the
alternate barrier model, discussed in this paper, back and forth
jumps are restricted to the nearest neighbor position. The reason
for these long-range backward correlations is quite intuitive: the
ions look for paths which can can be accessed rather easily. In
particular at low temperatures these paths can be rather extended.
Only for long times the ions manage to escape such a local path.
This picture is consistent with the percolation approach which has
been successfully applied to describe the dynamics in the random
barrier and the random energy model \cite{Dyre,Percolation}.

It may be interesting to compare this scenario with that of
supercooled liquids. There all particles move on the same time
scale. No pseudo-external fields are present. The complexity of
the dynamics is due to the necessity of cooperative dynamics of
the strongly interacting particles.  A simple picture is to view a
particle localized in the cage of the adjacent particles. The
relevant relaxation process is to escape this local cage and
afterwards being trapped in a new cage. In this case the first
moment $\bar{r}(r_{01})$ is constant for values of $r_{01}$ larger
than the typical nearest-neighbor distance \cite{Doliwa98}. This
is due to the fact that after leaving the initial cage no
significant memory to that cage is left. The $r_{01}$-dependence
of $\bar{r}(r_{01})$ for $r_{01} > d_0$ shows that this simple
cage picture cannot be used for the ionic dynamics. A possible
explanation for this effect is the relevance of the static
disordered energy landscape in ion conductors which leads to a
dramatic reduction of multi-particle correlations \cite{Heuer02}
such that cages, formed by adjacent ions, are less relevant.
Alternatively, one might argue that the presence of long-range
Coulomb interaction gives rise to long-range backward
correlations. This aspect still has to be clarified.

In the alternate barrier model the dependence of $\bar{r}(d_0)$ on
$t_{01}$ reflects the values of the lowest and highest relevant
barriers present in the system. On the time scale for which the
highest relevant barrier is crossed no back and forth dynamics
should be visible. Indeed, we see in \figref{fig_8}, e.g. for T = 750 K,
that the back and forth dynamics becomes small for $t_{01}$ of the
order of 1 ns. The relaxation time, i.e. the decay time of
$S(q_{max},t)$, is of the same order. Within the alternate barrier
model one expects that for $t_{01}$ somewhat smaller than
$1/\Gamma_2$ but still larger than $1/\Gamma_1$ one expects a
linear dependence of $\bar{r}(d_0)$ on $t_{01}$ in marked contrast
to the numerical results. This clearly shows that in contrast to
the alternate barrier model the lithium metasilicate system is
characterized by a broad distribution of relevant barriers.
Furthermore no short-time limit is visible in \figref{fig_8} but
$\bar{r}(d_0)$ decreases also for the shortest times which could
be analysed. This indicates the presence of transitions with very
small saddles or, possibly, broad anharmonic potentials which are
already relevant on the time scale of a few hundred femtoseconds.

The alternate barrier model differs from our lithium metasilicate
system also in another respect. For the alternate barrier model
the typical jump time is of the order of $1/\Gamma_1$. One can
show that (i) the decay of $w(t)$ is strongest for exactly this
time scale and (ii) $S(q=2\pi,t \approx 1/\Gamma_1)$ is already
significantly smaller than unity. For the lithium metasilicate
system at, e.g., $T = 640$ K the decay of $w(t)$ (due to
long-range processes; see above) is strongest in the ps range
whereas $S(q,t)$ is still close to the short-time plateau value.
This discrepancy could be alleviated by introducing an additional
variation of site energies. This would shift the decay of $S(q,t)$
to times of the order of $1/\Gamma_2$. On a qualitative level this
would imply that the short-time dynamics occurs in asymmetric
double-well potentials. It remains an important question whether
the same scenario also holds in lithium metasilicate, i.e. whether
fast back and forth jumps in strongly asymmetric double-well
potentials are present. Another scenario to rationalize the
above-mentioned discrepancy will be presented further below.

In Ref.\ \cite{Jund02} it has been reported that for sodium silicate at T = 2000 K
no backward correlations were observed. This is compatible with the
experimental data for sodium silicate where the dispersion disappears
for T $\approx$ 1000 K. [5]. For the present system the dispersion
disappears around a similar temperature. This shows up in a very little
time-dependence of $w(t)$ at 980 K in Fig.5 beyond 1 ps. Nevertheless,
even these weak backward correlations are directly visible by analysis
of the three-time correlations in Fig.8. As discussed below these
backward correlations may be due to a small number of ions. Thus a less
detailed analysis may oversee these backward correlations.

The approximate time-temperature superposition principle as seen
in \figref{fig_8} implies that the back-dragging effect remains the same if
analysed on appropriately adjusted time scales. Note that this
statement goes beyond the previous result that the mean square
displacement displays time-temperature superposition. As discussed
above the latter observable is only related to the
 $t_{01}\rightarrow 0$ limit of $\bar{r}(r_{01})$.
The time-temperature superposition in \figref{fig_8} may be used to
discriminate between different models of ion dynamics.

The significant heterogeneities, as characterized by the second
moment, is compatible with the above-mentioned broad distribution
of relevant barriers. Actually, such a distribution complicates
the interpretation of the dispersion $w(t \rightarrow
\infty)/w(0)$ in terms of correlated back and forth dynamics. This
is exemplified for the simple case of two (temporarily distinct)
ionic species with jump rates $\Gamma_1 \gg \Gamma_2$ which are
present with probabilities $p_1$ and $p_2$, respectively. If we
have in mind a log-gauss distribution of rates one should choose
$p_1 > p_2$. Generalizing Eq.\ref{firstmoment} we get
\begin{equation}
\frac{w(t \rightarrow \infty)}{w(t \rightarrow 0)} = \frac{1}{2} + \frac{1}{d_0}
\frac{p_1 \Gamma_1 \bar{r}^1 +p_2 \Gamma_2 \bar{r}^2}
{p_1 \Gamma_1 + p_2 \Gamma_2}
\end{equation}
The terms $\Gamma_i p_i$ imply that during the  short time
interval $t_{01}$ the probability of a particle of species i to
perform a hop is proportional to the rate and its occurrence
probability. In the limit, discussed above, and under the
additional assumption that $|\bar{r}^1|$ is not much smaller or
even larger than $|\bar{r}^2|$, this can be approximated by
\begin{equation}
\frac{w(t \rightarrow \infty)}{w(t \rightarrow 0)} \approx \frac{1}{2} +
\frac{\bar{r}^1}{d_0}.
\end{equation}
This result shows that the dispersion is to a large degree
determined by the backjump properties of the {\it fast} species.
Thus in case of significant dynamic heterogeneities one has to be
careful to relate the dispersion of the mean square displacement
to the average backjump properties (here: $(p_1 \bar{r}^1 + p_2
\bar{r}^2)/d_0$) of the ions.

This observation allows one to envisage another explanation of the
very different time scales where the decay of $w(t)$ is maximum
(ps regime) and where the incoherent scattering function
$S(q_{max},t)$ decays (ns regime); see above. A few fast ions with
significant back- and forth correlations can dominate the
time-dependence of $w(t)$ but hardly contribute to the decay of
the incoherent scattering function. Actually, for the random
energy model we have observed that the back and forth correlations
are strongest for the fast particles such that this effect is
indeed present there.  Whether it is this scenario or the
above-discussed presence of energetic disorder, which dominates
the different time dependence of $w(t)$ and $S(q_{max},t)$ for
lithium metasilicate, awaits further clarification.

We have shown that the degree of heterogeneity depends on
temperature. In contrast, the incoherent scattering function
$S(q_{max},t)$ fulfills the time-temperature superposition
principle. Actually, the same observations have been made for
glass-forming liquids \cite{Doliwa99}. For these systems the
non-exponentiality of $S(q_{max},t)$ mainly reflects the broad
distribution of relaxation times. Thus naively one would expect a
lower value of the KWW-exponent $\beta$ with increasing degree of
heterogeneity. It still has to be shown why at least in the
temperature range, accessible to simulations, this is not the
case.

It may be interesting to compare our approach with that of
Habasaki and Hiwatari \cite{Habasaki02}. For the same system at T
= 700 K they determined during 1 ns the  distribution of square
displacements of all particles. They observed a large variance,
thus indicating some distribution of relaxation times. On a
qualitative level this result is compatible with the results
reported above. One advantage of the present approach is that via
study of the $r_{01}$-dependence also information about length
scales are available. Furthermore, the value of $V(t \rightarrow
0)$ has a direct interpretation in terms of local rate
distributions.

Having identified several properties of the complex ion dynamics in
quantitative terms one would like to relate them to more
microscopic properties like the distribution of oxygens around the
lithium ions and to see whether mainly the interaction among the
different ions or the interaction with the basically static network
gives rise to the observations reported in this work.

\begin{acknowledgments}
In this work we have profited from helpful discussions with C.
Cramer, K. Funke, J. Habasaki, H. Lammert, and B. Roling.
\end{acknowledgments}

\appendix
\section{\label{app1}}

Calculation of $\bar{r}(r_{01}=1)$ proceeds in two steps. First we
calculate the probability that a particle has moved by one unit to
the right during some time $t_{01}$. Since we have two
non-equivalent sites we have to calculate, on the one hand, the
transition probability $q_{eo}$ ($eo$ stands for even-odd) from
$r=0$ to $r=1$ and, on the other hand, the transition probability
$q_{oe}$ from $r=1$ to $r=2$. In the second step we calculate the
average motion $\Delta r_e$ (starting, e.g., from $r=2$) and
$\Delta r_o$ (starting, e.g., from $r=1$) during the second time
interval. One expects $\Delta r_e > 0$ and $\Delta r_o < 0$ and
for reasons of symmetry  $\Delta r_e = - \Delta r_o$ which we
abbreviate as $\Delta r$. With this information we can finally
calculate
\begin{equation}
\label{eqrbar} \bar{r}(1) = \frac{\Delta r_o q_{eo} + \Delta r_e
q_{oe}}{ q_{eo} +  q_{oe}} = - \Delta r \frac{q_{eo} -
q_{oe}}{q_{eo} + q_{oe}}.
\end{equation}

For the calculation of $q_{eo}$ and $q_{oe}$ we take into account
that in the limit $t_{01} \Gamma_2 \ll 1$ multiple transitions
over the higher barrier can be neglected. We start with a particle
either at $r = 0$ or $r = 1$ and we are interested in the
probability to be at $r=1$ or $r=2$ after time $t_{01}$,
respectively. Neglecting those terms, which are only relevant in
case that a particle has crossed a high barrier at least twice, we
end up with the following system of rate equations for the site
populations $p_i$
\begin{eqnarray}
(d/dt) p_0 &  = & -(\Gamma_1 + \Gamma_2) p_0 + \Gamma_1 p_1 \\
(d/dt) p_1 &  = & -(\Gamma_1 + \Gamma_2) p_1 + \Gamma_1 p_0 \\
(d/dt) p_2 &  = & -\Gamma_1  p_2 + \Gamma_1 p_3 + \Gamma_2 p_1 \\
(d/dt) p_3 & = & -\Gamma_1 p_3 + \Gamma_1 p_2.
\end{eqnarray}
This set of differential equations can be directly solved.
$q_{eo}$ can be identified as $p_1$ with initial condition $r=0$
and $q_{oe}$  as $p_2$ with initial condition $r=1$. One obtains
after a short calculation
\begin{equation}
q_{eo} = (1/2)(1 - \Gamma_2 t_{01})(1 - \exp(-2\Gamma_1t_{01}))
\end{equation}
and
\begin{equation}
q_{oe} = \frac{\Gamma_2}{4 \Gamma_1} [1 - \exp(-2\Gamma_1t_{01}) +
\Gamma_1 t_{01} (1 + \exp(-2\Gamma_1t_{01})].
\end{equation}
Thus we finally get (using again $\Gamma_1 \gg \Gamma_2$)
\begin{equation}
\frac{q_{eo} - q_{oe}}{q_{eo} + q_{oe}} = 1 - 2
\frac{\Gamma_2}{\Gamma_1}  \, \, (t_{01} \Gamma \ll 1)
\end{equation}
and
\begin{equation}
\frac{q_{eo} - q_{oe}}{q_{eo} + q_{oe}} = 1 - \Gamma_2 t_{01} \,
\, (t_{01} \Gamma \gg 1).
\end{equation}

As a second ingredient we want to calculate $\Delta r$ which is
the average coordinate $\langle r \rangle$ after time $t_{12}$ for
the initial condition $r=0$. Using the standard trick of
introducing the functions
\begin{equation}
S_q \equiv \sum_j p_j \exp(iqj)
\end{equation}
and
\begin{equation}
T_q \equiv \sum_j (-1)^j p_j \exp(iqj)
\end{equation}
one can write down two linear differential equations involving
$T_q$ and $S_q$

\begin{eqnarray}
(d/dt)S_q(t) & = & -(\Gamma_1 + \Gamma_2) S_q (1 - \cos q) - i
(\Gamma_1 - \Gamma_2) T_q  \sin q \\ (d/dt) T_q(t) & = &
-(\Gamma_1 + \Gamma_2) T_q (1 + \cos q) + i (\Gamma_1 - \Gamma_2)
S_q \sin q
\end{eqnarray}

which can be solved with standard methods after specification of
the initial condition. Of interest for us is the expectation value
$\langle r \rangle$ for the initial condition $r(0) = 0$. This
expression can be calculated from
\begin{equation}
\langle r \rangle = -i \lim_{q \rightarrow 0} (d/dq) S_q.
\end{equation}

It turns out
\begin{equation}
\Delta r = (1/2) \frac{\Gamma_1 - \Gamma_2}{\Gamma_1 +
\Gamma_2}\left ( 1 - \exp(-2(\Gamma_1 + \Gamma_2) t_{12})  \right
)
\end{equation}
which in the limit $\Gamma_1 \gg \Gamma_2$ can be rewritten as
$\Delta r = (1/2)(1 - 2 \Gamma_2/\Gamma_1) (1 - \exp(-2\Gamma_1
t_{12})) $.

Thus we have calculated all ingredients which are necessary for
determination of $\bar{r}(1)$.

\section{\label{app2}}
Here we calculate the short-time  limit of the random trap model.
We always choose $t_{01} = t_{12} = t$. We define the probability
that a trap has the escape rate $\Gamma_i$ as $p_i$. In
equilibrium the probability that a particle is in a trap with
escape rate $\Gamma_i$ is proportional to $p_i/\Gamma_i$. Then we
can write
\begin{equation}
v(r_{01} = 1,t) = \frac{\sum_{i,j} (p_i/\Gamma_i) p_j q_{i}(t)
r^2_{j}(t)}{\sum_{i,j} (p_i/\Gamma_i) p_j q_{i}(t)}.
\end{equation}
Here $q_i(t)$ denotes the probability that after time $t$ the
particle has moved by one unit, starting in a trap with escape
rate $\Gamma_i$ and $r^2_j(t)$ the short-time expression for the
mean square displacement, starting from a trap with rate
$\Gamma_j$. One simply has $q_i(t) = \Gamma_i t$ and $r^2_j(t) =
\Gamma_j t$. After a short calculation one obtains
\begin{equation}
v(r_{01} = 1,t \rightarrow 0) = \sum_i p_i \Gamma_i \equiv \langle
\Gamma \rangle.
\end{equation}
In analogy one obtains for the short-time expansion of $v(r_{01} =
0,t)$
\begin{equation}
v(r_{01} = 0,t) = \frac{\sum_{i,j} (p_i/\Gamma_i) p_j (1-q_{i}(t))
r^2_{i}(t)}{\sum_{i,j} (p_i/\Gamma_i) p_j (1-q_{i}(t))} =
\frac{1}{\langle 1/\Gamma\rangle}.
\end{equation}
Thus one obtains
\begin{equation}
\frac{v(r_{01}=1,t)}{ v(r_{01} = 0,t)} = \langle \Gamma \rangle
\langle 1/\Gamma \rangle.
\end{equation}


\begin{thebibliography}{99}
\bibitem{expl1} One can imagine that for very short $t_{01}$ the only
chance to move $r_{01} = 0.5$ is to start at -0.5 and roll down to
the minimum, i.e. $r_1 = 0$.
\end{thebibliography}

\end{document}